\documentclass[12pt,preprint]{emulateapj}
\usepackage{natbib}
\usepackage{aas_macros}
\DeclareMathAlphabet{\mathsc}{OT1}{cmr}{m}{sc}

\begin{document}

\title{Dust Heating by Low-mass Stars in Massive Galaxies at $z<1$}

\author{ 
M. Kajisawa\altaffilmark{1,2},
T. Morishita\altaffilmark{3,5},
Y. Taniguchi\altaffilmark{2}, 
M. A. R. Kobayashi\altaffilmark{2},
T. Ichikawa\altaffilmark{3},
Y. Fukui\altaffilmark{4}
}


\altaffiltext{1}{Graduate School of Science and Engineering, Ehime University, 
        Bunkyo-cho, Matsuyama 790-8577, Japan
        {\it e-mail kajisawa@cosmos.phys.sci.ehime-u.ac.jp}}
\altaffiltext{2}{Research Center for Space and Cosmic Evolution, 
Ehime University, Bunkyo-cho, Matsuyama 790-8577, Japan}
\altaffiltext{3}{Astronomical Institute, Tohoku University, Aramaki, Aoba, Sendai 980-8578, Japan}
\altaffiltext{4}{Department of Astrophysics, Nagoya University, Chikusa-ku, Nagoya 464-8602, Japan}
\altaffiltext{5}{Institute for International Advanced Research and Education, Tohoku 
University, Aramaki, Aoba, Sendai 980-8578, Japan}

\shortauthors{M. Kajisawa et al.}
\shorttitle{Dust heating by low-mass stars in massive galaxies}

\begin{abstract}
Using the {\it Hubble Space Telescope}/Wide Field Camera 3 imaging data 
and multi-wavelength photometric catalog, we investigated the dust temperature 
of passively evolving and star-forming galaxies at 
$0.2<z<1.0$ in the CANDELS fields.
We estimated the stellar radiation field by low-mass stars from 
the stellar mass and surface brightness profile of these galaxies and then calculated 
their steady-state dust temperature.
At first, we tested our method using nearby early-type galaxies with the deep FIR  
data by the Herschel Virgo cluster survey 
and confirmed that the estimated dust temperatures are consistent with the observed 
temperatures within the uncertainty.  
We then applied the method to galaxies at $0.2<z<1.0$, and 
found that most of passively evolving galaxies with $M_{\rm star} > 10^{10} M_{\odot}$
have a relatively high dust temperature of $T_{\rm dust} > 20$ K, 
for which the formation efficiency of molecular hydrogen on the surface 
of dust grains in the diffuse ISM is expected 
to be very low from the laboratory experiments.
The fraction of passively evolving galaxies strongly depends on 
the expected dust temperature at all redshifts and increases rapidly with increasing 
the temperature around $T_{\rm dust} \sim 20$ K. 
These results suggest that the dust heating by low-mass stars in massive galaxies 
plays an important role for the continuation of their passive evolution, 
because the lack of the shielding effect of the molecular hydrogen 
on the UV radiation can prevent the gas cooling and formation of new stars.
\end{abstract}



\section{INTRODUCTION}

Galaxies are generally divided into two populations, namely, star-forming galaxies 
and passively evolving galaxies with little current star formation (e.g., \citealp{bel04}). 
At redshift $z<1$, most of massive galaxies are passively evolving, 
while low-mass galaxies tend to actively form new stars (e.g., \citealp{kau03}; 
\citealp{bun06}). 
Then massive galaxies are considered to have finished their star formation 
in the early universe, and the star formation can be seen only in gradually 
less massive galaxies as time passes, the so-called down-sizing evolution of galaxies
\citep{cow96}. 
Since cold gas is expected to successively accrete onto massive galaxies 
in the structure formation model of the $\Lambda$CDM universe, such down-sizing effect 
suggests that some quenching mechanisms for the star formation continue to work 
in massive galaxies at $z < 1$. As a possible such mechanism, 
“radio-mode” feedback, in which the kinetic energy of radio jets 
from active galactic nuclei heats the gas in the dark matter halo and 
prevents its radiative cooling, has been extensively discussed in many studies (e.g., 
\citealp{fab12}).

On the other hand, several observational studies reported that the surface 
stellar mass density of galaxies is better correlated with their star formation activity 
and color than stellar mass itself at both high and low redshifts (e.g., \citealp{kau06}; 
\citealp{fra08}; \citealp{wil10}). 
In fact, passively evolving massive elliptical galaxies tend to have higher 
surface mass density than star-forming spiral galaxies in the present universe (e.g., 
\citealp{she03}). 
The surface stellar mass density is considered to be related closely 
with the volume number density of stars within the galaxy, 
in particular that of long-lived low-mass stars which dominate 
the total stellar mass of the galaxy.
In this study, we relate the surface stellar mass density to the equilibrium temperature of 
dust grains heated by low-mass stars in galaxies, and discuss its relationship with 
their star formation activity. 
The formation of molecular hydrogen is considered to occur 
on the surface of dust grains (e.g., \citealp{gou63}), 
and the dust temperature can affect its formation in the diffuse 
interstellar medium (e.g., \citealp{kat99}).
If the dust grains have been heated up to sufficiently high temperatures in massive galaxies,  
the formation of molecular hydrogen would continue to be suppressed for a long time, 
and the lack of the shielding effect of the molecular hydrogen on the UV radiation
 would prevent the gas cooling and formation of new stars.  
In order to examine whether the stellar radiation field by low-mass stars 
is sufficiently strong or not, we calculate the stellar radiation field from stellar mass  
and surface brightness profile and then estimate the steady-state dust temperature 
for galaxies at $0.2 < z < 1.0$ in the CANDELS survey fields \citep{gro11}.

Section 2 describes the sample selection and the method to estimate the stellar radiation 
field and the dust temperature. In Section 3, we check our method by comparing with 
 the direct measurements of the dust temperature for nearby early-type galaxies 
by the Herschel satellite. 
We present the results for our main sample of 
galaxies at $0.2<z<1.0$ and discuss them in Section 4.  
Throughout this paper, magnitudes are given in the AB system.
 We adopt a flat universe with $\Omega_{\rm matter}=0.3$, $\Omega_{\Lambda}=0.7$, 
and $H_{0}=70$ km s$^{-1}$ Mpc$^{-1}$.

\section{SAMPLE AND ANALYSIS}
In this study, we used the 3D-HST WFC3-selected photometric catalog 
version 4.1\footnote{http://3dhst.research.yale.edu/Data.php} and the WFC3 
imaging data for the five CANDELS/3D-HST fields (GOODS-North and South, AEGIS, COSMOS, and 
UKIDSS UDS) released by the 3D-HST team \citep{ske14}. 
The source detection was performed on the combined WFC3 images made from the 
$J_{\rm F125W}$, $JH_{\rm F140W}$, and $H_{\rm F160W}$-bands data.
The 3D-HST team performed the SED fitting and estimated the photometric redshift, 
rest-frame colors, and stellar mass using the photometric data from $U$-band to 
$Spitzer$/IRAC 8.0 $\mu$m band. 
Chabrier Initial Mass Function (IMF) \citep{cha03} 
was assumed in their calculation of stellar mass of galaxies. 
At first, we selected extended sources with $H_{\rm F160W} < 25$ mag and ``use\_flag'' $=$ 1, 
which ensures a quality of the photometry and SED fitting procedure \citep{ske14}, 
from the catalog to achieve a high completeness 
for the surface brightness profile fitting with the WFC3 images described below. 
We then made a sample of galaxies with $M_{\rm star} > 10^{9} M_{\odot}$ at 
$0.2<z_{\rm phot}<1.0$. 
The magnitude limit of $H_{\rm F160W} < 25$ is sufficient to sample galaxies with 
$M_{\rm star} > 10^{9} M_{\odot}$ up to $z\sim1$.


\begin{figure}
\begin{center}
\includegraphics[width=83mm]{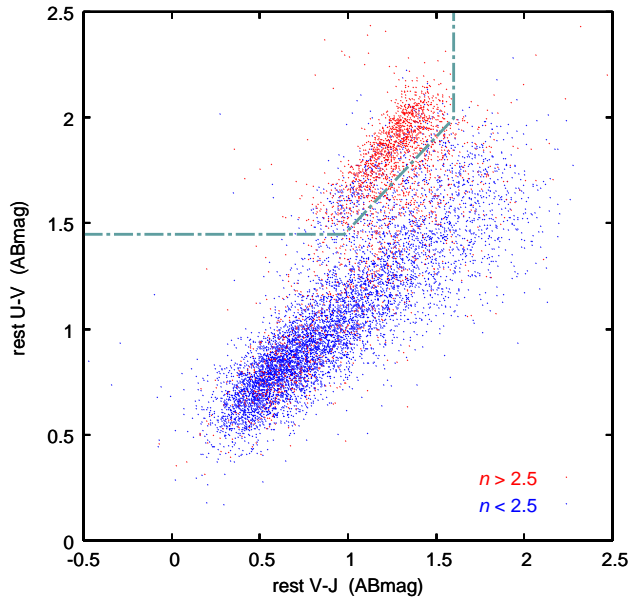}
\caption{
The rest-frame $U-V$ vs. $V-J$ two-color diagram for galaxies with 
$M_{\rm star} > 10^{9} M_{\odot}$ at $0.2<z<1.0$ in the CANDELS fields.
The dotted-dashed line shows the color criteria for the passively evolving galaxies.
Red dots show galaxies with a S{\'e}rsic index of $n > 2.5$ and 
blue ones represent galaxies with $n < 2.5$.
}
\label{fig:uvj}
\end{center}
\end{figure}
We divided the sample galaxies into the passively evolving and star-forming populations 
on the rest-frame $U-V$ vs. $V-J$ two-color diagram as seen in Figure 
\ref{fig:uvj}, following \citet{wil09}.
The color selection criteria for passively evolving galaxies we used are as follows, 
\begin{equation}
U-V>0.88\times(V-J)+0.59\ \&\ V-J<1.6\ \&\ U-V>1.4.
\end{equation}
The other galaxies are classified as star-forming galaxies. 
We applied this selection for galaxies at $0.2 < z < 1.0$, 
while \citet{wil09} changed the criteria slightly depending on redshift. 
We also used the slightly more stringent criterion of the $U-V$ 
color to prevent the contamination from star-forming galaxies. 
We confirmed that the small differences in the color criteria 
do not affect our results at all.

In order to estimate the dust temperature from the stellar radiation field by low-mass stars,  
we analyzed the surface brightness 
of our sample galaxies by ourselves, using the WFC3 $H_{\rm F160W}$-band images.  
The $H_{\rm F160W}$ band samples the rest-frame 8000--13000 \AA\ light for galaxies 
at $0.2<z<1.0$. 
We are interested in the radiation field made by low-mass stars, 
whose light dominates the near-infrared radiation. Therefore  
the $H_{\rm F160W}$-band data are suitable for our purpose.
The morphological K-correction within these rest-frame 
wavelengths is expected to be very small and it 
does not affect the estimate of the dust temperature as shown in the next section. 
Following our previous study \citep{mor14}, 
we used the publicly available code GALFIT \citep{pen02} to fit 
the surface brightness profiles 
of galaxies with the S{\'e}rsic profile \citep{ser63}. 
GALFIT provided us the best-fit S{\'e}rsic index $n$, half-light semi-major radius 
$a_{\rm e}$, and half-light semi-minor radius $b_{\rm e}$. 
As shown in previous studies at $z\lesssim 1$, 
galaxies with $n > 2.5$ tend to be passively evolving, 
while most galaxies with $n < 2.5$ are star-forming (Figure \ref{fig:uvj}). 
In total, we used 9281 galaxies at $0.2 < z < 1.0$ with the measured surface 
brightness parameters over a total area of 896.3 arcmin$^2$.
Out of 9281 sample galaxies, 2409 galaxies have $n>2.5$, while there are 
6872 galaxies with $n<2.5$.

\begin{figure}
\begin{center}
\includegraphics[width=83mm]{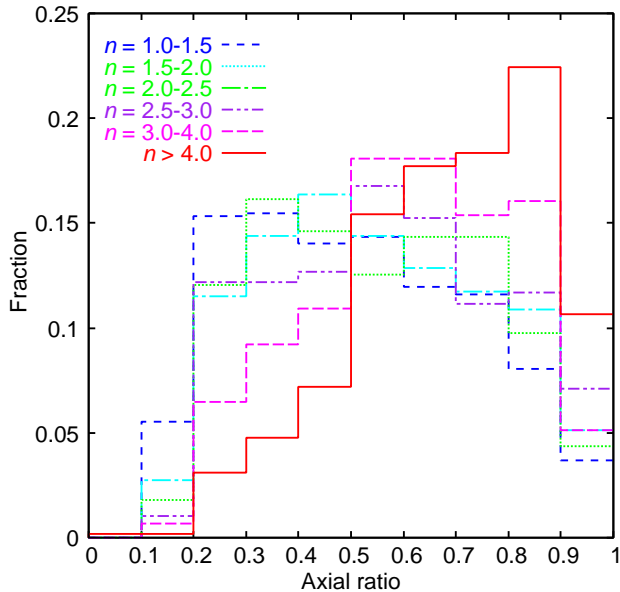}
\caption{
Distribution of the observed axial ratio $b_{\rm e}/a_{\rm e}$ for our sample galaxies  
with the different S{\'e}rsic indices. 
}
\label{fig:ar}
\end{center}
\end{figure}

We inferred the 3-dimensional distribution 
of stellar mass from the 2-dimensional surface brightness profile mentioned above, 
 in order to estimate the stellar radiation field. 
For simplicity, we assume a constant $M_{\rm star}/L_{\rm F160W}$ ratio within a galaxy, 
which means that the profile of the surface stellar mass density has the same shape as 
the $H_{\rm F160W}$-band surface brightness profile.
Under this assumption, 
we can slightly underestimate the stellar mass of the bulge and overestimate 
that of the disk for galaxies with a color gradient 
such as disk galaxies with red bulge and blue disk.
As a result, we can underestimate the dust temperature at a small radius
of the galaxy and overestimate at a large radius. However, the effect is relatively small 
because of the relatively small variance of the stellar M/L ratio in the rest-frame 
$\sim $ 10000 \AA, and in fact our results do not change significantly 
if we measure the dust temperature at the different radii such as $0.5 \times r_{\rm e}$ 
and $2.0 \times r_{\rm e}$ instead of $r_{\rm e}$. 
From the distribution of the axial ratio $b_{\rm e}/a_{\rm e}$ seen in Figure \ref{fig:ar}, 
we expect that most galaxies with $n > 2.5$ have spheroidal shapes, while galaxies 
with $n < 2.5$ tend to have thin disk shapes. 
For the distribution of stars in galaxies with $n > 2.5$, we used a spherically 
symmetric Hernquist-like profile \citep{her90},
\begin{equation}
\rho(r) = \frac{A}{r_{\rm c}^3}\frac{1}{(r/r_{\rm c})(1+r/r_{\rm e})^{\gamma}} , 
\end{equation}
where $r_{\rm c}$ and $\gamma$ are fitted so that the 2-dimensional projection of 
this profile is matched to the circularized S{\'e}rsic profile with the observed 
S{\'e}rsic index $n$ and half-light radius $r_{\rm e} = (a_{\rm e}b_{\rm e})^{1/2}$. 
$A$ is normalization constant determined by requiring that the integration of 
the profile is equal to the total stellar mass of each galaxy. 
For galaxies with $n < 2.5$, we used the following thin disk profile, 
\begin{equation}
\rho(r,\zeta) = A\exp\left [-b_n\left (\frac{r}{a_{\rm e}}\right )^{1/n}\right ]\exp\left [-b_l\left (\frac{\zeta}{0.1a_{\rm e}}\right )\right ].
\end{equation}
The profile along the disk radius is the observed S{\'e}rsic profile and that along the 
disk height is assumed to be the exponential profile. 
Note that we adopted the semi-major axis $a_{\rm e}$ as a radius of the disk 
to take account of the inclination effect. 
We also assumed a scale height of $0.1 \times a_{\rm e}$, which is a typical value 
for local disk galaxies. 
Even if we adopt $0.05 a_{\rm e}$ or $0.2 a_{\rm e}$ 
as a scale height instead of $0.1 a_{\rm e}$, 
the estimated dust temperature changes only by $\sim$ 2--3 \% (i.e., 0.3--1.0 K). 

We then calculated the stellar radiation field as 
\begin{equation}
U(r) = \left (\frac{M_{\rm star}}{L_{\rm bol}}\right )^{-1}\int\rho(r_{0})\frac{1}{4\pi|r-r_{0}|^{2}c}\ dV
\end{equation}
and
\begin{equation}
U(r,0) = \left (\frac{M_{\rm star}}{L_{\rm bol}}\right )^{-1}\int\rho(r_{0},\zeta_{0})\frac{1}{4\pi|r-r_{0}|^{2}c}\ dV
\end{equation}
for galaxies with $n > 2.5$ and $n < 2.5$, respectively. 
For the thin disk model, the radiation field at various radii on the disk plane 
(i.e., at $\zeta = 0$) was calculated. 
We here used the bolometric stellar mass-to-luminosity 
ratio $M_{\rm star}/L_{\rm bol}$ of a single 3 Gyr burst model with an age of 6 Gyr 
from the GALAXEV population synthesis library \citep{bru03} 
in order to consider the radiation from long-lived, low-mass stars. 
As shown in the next section, 
a change of the model age or duration of the star formation does not significantly 
affect our results as long as more than 1 Gyr passed since the star formation stopped.
While our sample galaxies are expected to have various star formation histories and 
stellar ages, we simply use the single 3 Gyr burst model with an age of 6 Gyr,  
because we aim to investigate how the dust temperature is maintained only 
by low-mass stars after the star formation stopped rather than estimate 
the exact temperature of galaxies at the observed epoch.
Therefore, for star-forming galaxies, we infer the dust temperature 
in the case that the star formation was quenched by some mechanism such as supernova feedback 
immediately after the observed epoch and then short-lived massive stars died.  
In other words, we aim to confirm whether the dust temperature continues to be high enough to 
suppress the production of the molecular hydrogen and to maintain the passive evolution 
 if the star formation has once stopped in the star-forming galaxies. 

Following \citet{gro12}, we simply adopted the following assumptions 
to estimate the steady-state dust temperature.
(1) The dust emission is expressed by the 
modified black body. (2) The stellar radiation field is not significantly affected 
by the dust absorption (i.e., optically thin). And, (3) the cooling by 
the dust emission balances the heating by the stellar radiation. 
Although star-forming galaxies are expected to have some amount of dust extinction, 
we here consider the contribution only from low-mass stars, which radiate their energy 
mainly at the NIR wavelength, and estimate the dust temperature 
expected after massive stars died for these galaxies as mentioned above. 
Therefore the assumption (2) is reasonable for our purpose.  
Under these assumptions, the dust temperature is expressed as follows \citep{dra11},
\begin{eqnarray}
\lefteqn{T_{\rm dust} =}\nonumber \\
&\left (\frac{h\nu_{0}}{k_{\rm B}}\right )^{\beta/(4+\beta)}\left [\frac{\pi^4c}{60 \Gamma(4+\beta) \zeta(4+\beta) \sigma}\frac{\left <Q_{\rm abs}\right >_{*}}{Q_{0}}\right ]^{1/(4+\beta)}U^{1/(4+\beta)},
\end{eqnarray}
where $h$, $k_{\rm B}$, and $\sigma$ are the Planck constant, Boltzmann constant, 
and Stephen-Boltzmann constant, respectively. 
$Q_{0}$ is the dust absorption cross-section at a reference wavelength of 
$\lambda_{0} = 100$ $\mu$m and $\left <Q_{\rm abs}\right >_{*}$  
is the spectrum-averaged absorption cross-section. 
The emissivity slope of $\beta = 2$ and the Milky Way dust model by \citet{wei01} 
with $R_{V} = 3.1$ were assumed.
We calculated $\left <Q_{\rm abs}\right >_{*}/Q_{0} \sim 209$ from the Milky Way dust model 
and the stellar SED of the single 3 Gyr burst model with an age of 6 Gyr. 
Since the stellar radiation field becomes stronger at smaller radii in a galaxy, 
the estimated dust temperature increases with decreasing radius. 
In order to ensure the dust temperature is higher than a given value over 
a significant volume of the galaxy, we adopt the temperature at a radius of 
$r = r_{\rm e}$. 
Even if we use the temperature at a different radius such as $0.5 \times r_{\rm e}$ or 
$2.0 \times r_{\rm e}$,  
the distribution of the dust temperature only slightly shifts to 
higher or lower values as shown in the next section 
and the results in this paper do not change.

\begin{figure}
\begin{center}
\includegraphics[width=83mm]{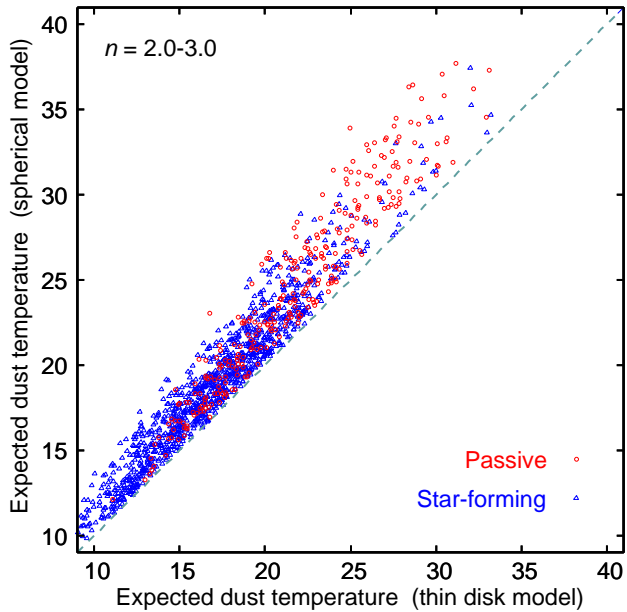}
\caption{Comparison between the expected dust temperatures estimated with 
the thin disk model and the spherically symmetric model of the 3-dimensional 
distribution of stars for galaxies with $n = $ 2.0--3.0. 
}
\label{fig:n2030}
\end{center}
\end{figure}
We also checked the effects of our simplification for the 3-dimensional distribution 
of stellar mass by comparing the dust temperatures estimated with the 
spherically symmetric model and the thin disk model of the same objects. 
In Figure \ref{fig:n2030}, we compare the dust temperatures calculated with 
the spherical model and the thin disk model for galaxies with $n = $ 2.0--3.0. 
Our spherical model predicts slightly higher temperatures than the thin disk model, 
and the mean and dispersion of the differences between these models 
are $\Delta T = 1.9 \pm 1.2$ K. The difference depends on the axial ratio, and 
the difference is very small ($\Delta T \lesssim 1$ K) for objects with 
$b_{\rm e}/a_{\rm e} \sim 1$, while it becomes larger 
($\Delta T \sim 4$ K) for those with $b_{\rm e}/a_{\rm e} \sim $ 0.2--0.4.
Since the differences between the two extreme cases (i.e., spherically symmetry and 
thin disk) is relatively small, we expect that our simplification for the distribution 
of stars does not significantly affect our results.
In fact, we obtain nearly the same results if we divide galaxies 
into the spheroidal shapes and thin disks at $n = 2.0$ or 3.0 instead of $n = 2.5$.

\begin{figure}
\begin{center}
\includegraphics[width=83mm]{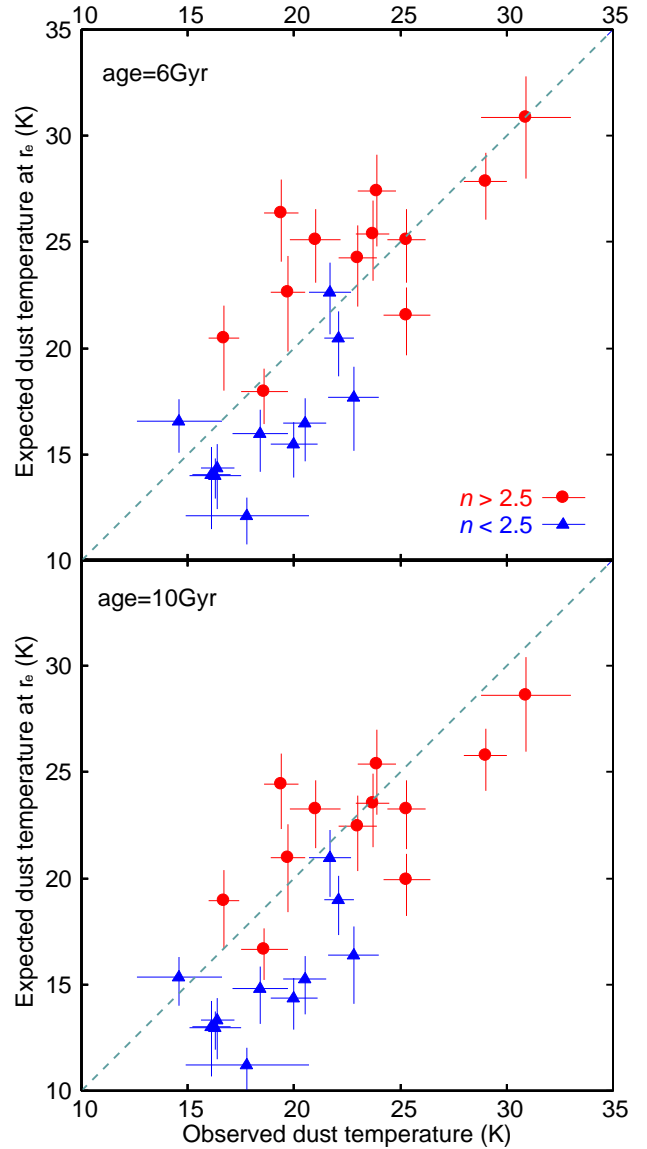}
\caption{
Comparison between the observed dust temperature from \citet{dis13} and 
the expected temperature which we estimated
 from the stellar mass and surface brightness profile for early-type galaxies 
in the Virgo cluster. The upper (bottom) panel shows the expected temperature 
for which we used the SED model of the single 3 Gyr burst model with an age of 
6 Gyr (10 Gyr) in the calculation. 
Circles show galaxies with the S{\'e}rsic index of $n > 2.5$, 
while triangles represent those with $n<2.5$.
The errors in the expected temperatures are based on the uncertainty in stellar 
mass from \citet{dis13} and that in the radius for which we assume 
30 \% error taking account of the systematic uncertainty due to the errors in the  
background estimate.   
}
\label{fig:vcc}
\end{center}
\end{figure}
\section{COMPARISON WITH DIRECT MEASUREMENTS}
In this section, we check the method described in the previous section,
 using the direct measurements of the dust temperature of local early-type galaxies.
We used the measurements by \citet{dis13} for early-type galaxies in the Virgo cluster.
\citet{dis13} fitted the Herschel 100, 160, 250, 350, and 500 $\mu$m-bands photometry 
from the Herschel Virgo cluster survey \citep{dav12}  
with the modified black body with $\beta = 2$ 
to measure the temperature of cold dust. Their deep 5-bands data from 100 to 500 $\mu$m 
enable the estimation of the dust temperature with relatively high accuracy. 
Since their sample has a wide range of stellar mass, we can expect 
a wide range of stellar surface density and dust temperature, which is essential for 
our purpose, although the sample is limited to the cluster galaxies. 
\citet{dis13} also estimated the stellar mass of the sample galaxies from their optical-NIR 
colors. The estimate of the stellar M/L ratio was based on the population synthesis model 
by \citet{bru03} and the Chabrier IMF was assumed. 
We used the dust temperature and stellar mass from Table 4 of \citet{dis13}.   

In order to measure the surface brightness profile of these galaxies, 
we used $z$-band images from the Sloan Digital Sky Survey (SDSS; 
\citealp{yor00}; \citealp{ahn14}). 
These data have a pixel scale of 0.4 arcsec/pixel and the PSF FWHM of $\sim$ 1.5 arcsec. 
We performed the same procedure of the surface brightness fitting with GALFIT to 
determine the best-fit S{\'e}rsic index $n$, half-light semi-major radius 
$a_{\rm e}$, and half-light semi-minor radius $b_{\rm e}$. 
Using the fitted parameters of the S{\'e}rsic profile from the SDSS $z$-band data 
and the stellar mass from \citet{dis13}, we then estimated the dust temperature at 
a radius of $r = r_{\rm e}$ as described in the previous section.
We excluded the objects for which the dust temperature cannot be determined by 
the FIR SED fitting of \citet{dis13} or the effective radius cannot be reliably 
derived in the surface brightness fitting, for example, 
due to their faintness on the SDSS images.

The upper panel of Figure \ref{fig:vcc} shows a comparison between the observed 
dust temperature from \citet{dis13} and the expected dust temperature we derived from 
the stellar mass and surface brightness profile. It is seen that the expected dust 
temperature correlates relatively well with the temperature estimated from the 
FIR observations, although there is some dispersion. The mean and dispersion of  
the differences between the observed and expected temperatures 
are $\Delta T = -0.4 \pm 3.3$ K. 
We also estimated the dust temperature using 
the single 3 Gyr burst model with an age of 10 Gyr as the star formation history 
instead of that with an age of 6 Gyr, because the 10 Gyr age can be more suitable for 
these early-type galaxies in the cluster at $z\sim0$. The result is shown in 
the bottom panel of Figure \ref{fig:vcc}. The expected temperatures become slightly 
lower values than those in the case with the 6 Gyr age.
In this case, the difference between the observed and expected temperatures 
is $\Delta T = -1.9 \pm 3.1$ K. The mean difference between the temperatures estimated 
from the model SEDs with the 6 Gyr and 10 Gyr ages is $\sim 1.5$ K, which is smaller 
than the dispersion around the mean. 
As mentioned in the previous section, the difference 
in the assumed model age does not affect so strongly the estimated dust temperature. 

We also checked the expected dust temperatures at the different radii, namely 
$r = 0.5\times r_{\rm e}$ and $r = 2.0\times r_{\rm e}$ in the galaxies. 
The differences between the observed temperature and the expected ones with the 6 Gyr
 age model are $\Delta T = 2.8 \pm 4.1$ K for $r = 0.5\times r_{\rm e}$ and 
$\Delta T = -3.9 \pm 2.8$ K for $r = 2.0\times r_{\rm e}$. 
The expected dust temperature becomes higher or lower by $\sim$ 3--4 K on average, 
when we use the temperature at $r = 0.5\times r_{\rm e}$ or $r = 2.0\times r_{\rm e}$. 
The results in the next section do not change even if we use the different radii.
The expected dust temperature at $r = r_{\rm e}$ seems to show a better agreement 
with the observed temperature than those at the smaller and larger radii, which 
may be consistent with the relatively compact morphology of the FIR dust emission 
in these galaxies \citep{dis13}. 

Furthermore, we performed the same analysis with $J$-band images from the 2-Micron All Sky 
Survey (2MASS; \citealp{skr06}) in order to 
examine the effect of the morphological K-correction on the estimate of the dust temperature.
The pixel scale of these data is 1.0 arcsec/pixel and the PSF FWHM is $\sim$ 3 arcsec.
The difference between the observed and expected temperatures is $\Delta T = 1.5 \pm 
4.1$ K. The difference between those with the $z$-band and $J$-band images is 
$\sim$ 2 K on average, which is considered to be relatively small 
if we take account of the dispersion around 
the mean and the different depth and spatial resolution between the SDSS $z$-band and 
2MASS $J$-band data.

In summary, the dust temperature estimated from the stellar mass and surface brightness 
profile agrees relatively well with the observed temperature for early-type 
galaxies in the Virgo cluster with a dispersion of $\sim$ 3--4 K.
There can be also the systematics of $\sim$ 3--4 K in the estimated temperature, 
depending on the assumed star formation history in the SED model,  
the wavelength where the surface brightness is measured, 
 and the radius used in the calculation of the dust temperature. 
Keeping in mind these uncertainties, we discuss the expected dust temperature for 
passively evolving and star-forming galaxies at $0.2<z<1.0$ in the following section.

\section{RESULT AND DISCUSSION}
\begin{figure}
\begin{center}
\includegraphics[width=83mm]{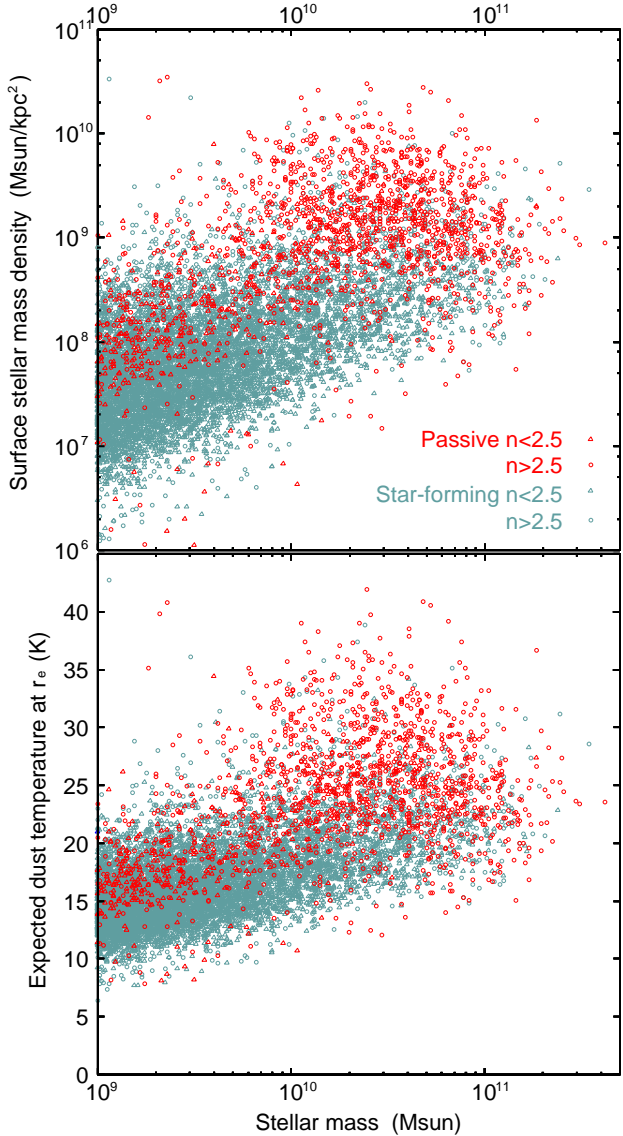}
\caption{
Distribution of surface stellar mass density (upper panel) and the expected 
dust temperature at a radius of $r = 2.0\times r_{\rm e}$ 
(lower panel) of galaxies at $0.2 < z < 1.0$ 
in the five CANDELS fields as a function of stellar mass. 
Color of the symbols shows passively evolving (red) and star-forming (light blue) 
populations. 
Circles show galaxies with the S{\'e}rsic index of $n > 2.5$, while triangles represent 
those with $n < 2.5$. The surface stellar mass density is calculated by 
$M_{\rm star}/2\pi r_{\rm e}^2$, which corresponds to the averaged value within 
a half-light radius.
}
\label{fig:smdt}
\end{center}
\end{figure}

We show the distribution of the surface stellar mass density and expected dust temperature 
as a function of stellar mass for passively evolving and star-forming galaxies 
at $0.2 < z < 1.0$ in Figure \ref{fig:smdt}. 
As previously reported (e.g, \citealp{kau06}; \citealp{fra08}; \citealp{wil10}), 
most galaxies with a high surface stellar mass density of  
$\Sigma_{\rm star} \gtrsim 10^{9} M_{\odot}$/kpc$^{2}$ are passively evolving ones 
especially at $M_{\rm star} > 10^{10} M_{\odot}$.
Similarly, most of passively evolving galaxies show a relatively high dust temperature 
of $T_{\rm dust} > 20$ K at $M_{\rm star} > 10^{10} M_{\odot}$. 
The expected dust temperature at smaller radii is higher 
because the stellar radiation increases with decreasing radius. 
Since we adopted the temperature at a radius of $r = r_{\rm e}$ in the figure, 
the dust temperature is expected to be higher than 20 K over a significant volume 
in most of massive passively evolving galaxies.

Since the energy density by the radiation from a star with a luminosity $L$ at a distance 
of $r$ is expressed by $L/4\pi r^{2}c$, 
the stellar radiation field by all stars at a given point in the galaxy is 
proportional to $M_{\rm star}/R^2$, where $R$ is the size of the galaxy. 
Therefore, the stellar radiation field in galaxies is 
directly related with their surface stellar mass density, and  
the close relationship between the dust temperature and 
the surface stellar mass density is naturally expected.

\begin{figure}
\begin{center}
\includegraphics[width=82mm]{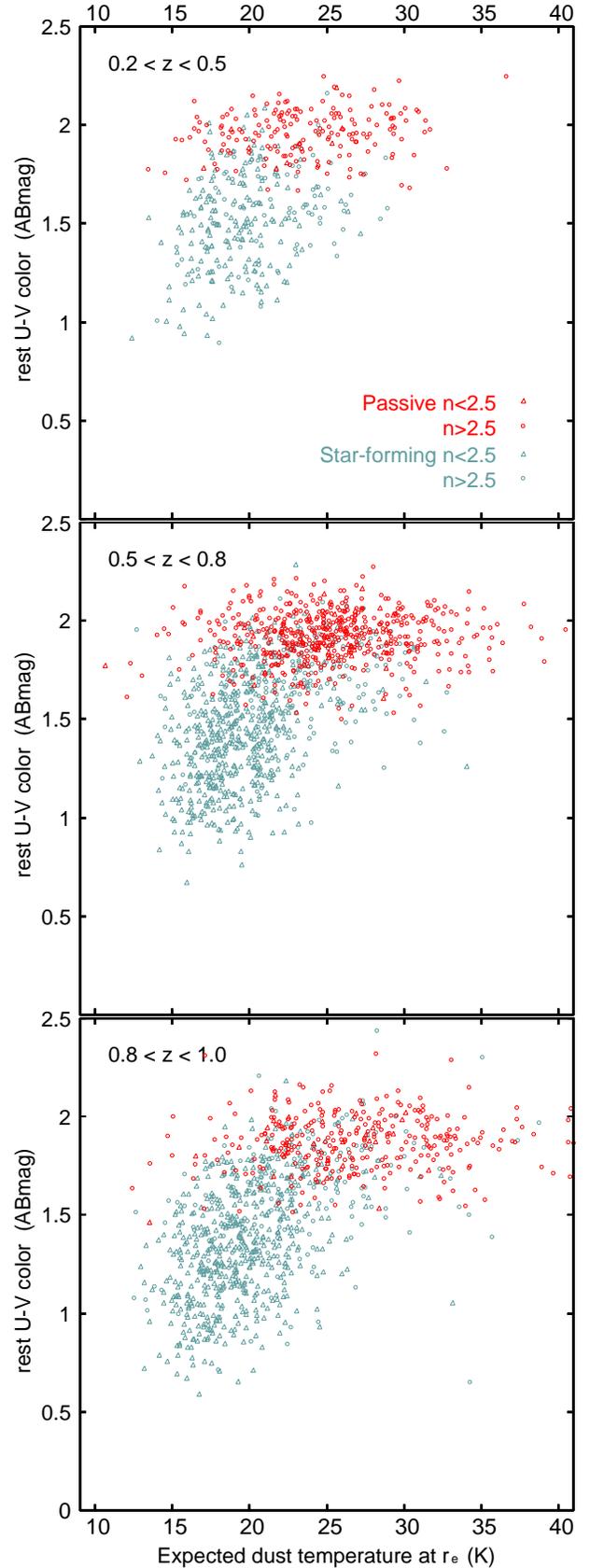} 
\caption{
The rest-frame $U-V$ color of galaxies with $M_{\rm star} > 10^{10} M_{\odot}$ 
as a function of the expected dust temperature for the different redshift bins.
Symbols are the same as Figure \ref{fig:smdt}.
}
\label{fig:uvt}
\end{center}
\end{figure}
\begin{figure}
\begin{center}
\includegraphics[width=83mm]{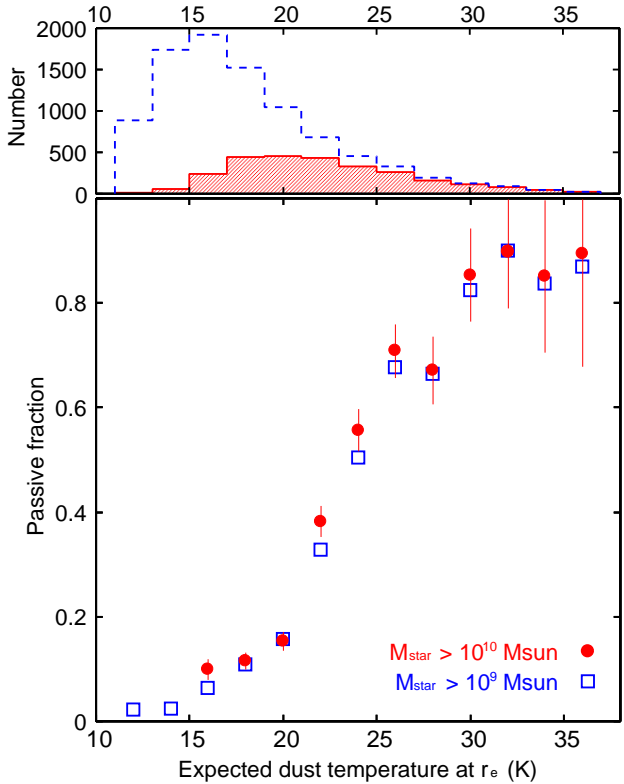} 
\caption{
Fraction of passively evolving galaxies in galaxies with $M_{\rm star} > 10^{10} M_{\odot}$ 
(solid circle) and $M_{\rm star} > 10^{9} M_{\odot}$ (open square) 
at $0.2 < z < 1.0$ as a function of the expected dust temperature. 
Data points for the temperature bins which include more than 10 passively evolving 
galaxies are plotted. The total number of galaxies with 
$M_{\rm star} > 10^{10} M_{\odot}$ 
(shaded histogram) and $M_{\rm star} > 10^{9} M_{\odot}$ (open histogram) is also shown 
in the upper panel.
}
\label{fig:fpt}
\end{center}
\end{figure}

Figure \ref{fig:uvt} shows the rest-frame $U-V$ 
color as a function of the expected dust temperature 
for galaxies with $M_{\rm star} > 10^{10} M_{\odot}$ in the different redshift bins. 
The rest-frame color becomes redder with increasing the dust temperature at all redshifts, 
and passively evolving galaxies on the red sequence dominate at $T_{\rm dust} \gtrsim $ 
20--25 K. 
The transition temperature between star-forming and passively evolving populations 
does not seem to significantly 
evolve with time. We show the fraction of passively evolving galaxies 
as a function of the dust temperature in Figure \ref{fig:fpt}. The fraction strongly 
depends on the dust temperature and increases rapidly with increasing 
the temperature around $T_{\rm dust} \sim 20$ K.

Recent observations of nearby early-type galaxies by the Herschel satellite revealed that 
massive early-type galaxies in fact tend to have a relatively high dust temperature of 
$T_{\rm dust} > 20$ K (e.g., \citealp{smi12}; \citealp{aul13}; \citealp{dis13}; 
\citealp{amb14}). 
\citet{smi12} and \citet{aul13} reported that early-type galaxies show systematically 
higher dust temperatures than late-type galaxies. \citet{dis13} suggested that massive 
early-type galaxies tend to have higher dust temperature than low-mass early-type ones.  
\citet{gro12} also reported that the bulge of M31, 
which has no young stellar population, shows a high dust temperature of 20-35 K, 
while the temperature in the star-forming disk is $\sim 17$ K. They found that 
the temperature profile of the M31 bulge can be explained by the stellar radiation 
field from its old stars. These studies are considered to be consistent with our results.

The dust temperature $T_{\rm dust} \sim 20 $ K corresponds to 
a temperature above which the formation 
efficiency of the molecular hydrogen in the diffuse ISM is expected 
to be very low from the laboratory experiments (e.g., \citealp{pir99}; \citealp{kat99}; 
\citealp{per05}; \citealp{lep09}). 
The molecular hydrogen cannot be efficiently formed in the gas phase, 
and its formation is considered to occur on the surface of dust grains, 
which work as catalysts \citep{gou63}.
However, the hydrogen atom that sticks to the surface of the grain quickly desorbs 
and cannot remain on the surface to encounter another atom and form a molecule at 
$T_{\rm dust} \gtrsim 20 $ K, although this may not be the case for more dense 
photo-dissociation regions of star-forming galaxies (e.g., \citealp{leb12}; \citealp{tie13}).
Thus if low-mass stars in massive galaxies has heated up the dust grains to 
$T_{\rm dust} > 20 $ K, 
the formation of the molecular hydrogen could have been suppressed in the diffuse ISM 
of these galaxies. 

Such a suppression of the formation of the molecular hydrogen is expected 
to play an important role  
for the passive evolution of massive galaxies. 
If we consider that the gas is accreting and cooling from higher temperatures 
after the star formation has once been stopped by, for example, intense supernova feedback, 
the HI gas is expected to settle to the equilibrium temperature 
(e.g., $T_{\rm gas} \sim 70$ K for the cold neutral medium in the Milky Way; \citealp{hei03})
 determined from a balance between the heating mainly 
by photoelectron from dust grains and cosmic-ray ionization 
and the radiative cooling by the emission lines such as [C{\footnotesize II}] 158 $\mu$m 
and [O{\footnotesize I}] 63 $\mu$m
\citep{dra11}.
If the formation of the molecular hydrogen proceeds in the cold HI gas, 
the self-shielding effect of the molecular hydrogen on the UV radiation 
would suppress the heating, which leads to further cooling of the gas and 
the star formation. 
On the other hand, if the formation of the molecular hydrogen in the diffuse ISM is 
suppressed due to the high dust temperature, the HI gas would keep the equilibrium 
temperature. 
Therefore the dust heating by long-lived, low-mass stars in massive galaxies 
can cause the continuation of their passive evolution.
Recent observational studies of the HI gas in local early-type galaxies 
found that a significant fraction of massive early-type galaxies have a moderate amount 
of HI gas (e.g., $10^{8}$--$10^{10} M_{\odot}$; \citealp{ser12}). 
The low star-formation activity of such galaxies may be related with their high dust 
temperature. 

On the other hand, several studies with radio observations also detected the CO lines 
in a significant fraction of nearby early-type galaxies 
(e.g., \citealp{wel10}; \citealp{you11}).
The morphology of the detected CO lines tends to be centrally concentrated and 
those early-type galaxies with the CO detection show the star formation activity at 
their central region (\citealp{ser12}; \citealp{you14}).
In the case that the cold gas can be concentrated into the center of the galaxy,  
for example, by the galaxy interactions, the molecular hydrogen may be formed 
and star formation may occur irrespective of the dust temperature, because the high 
density of the gas and dust can lead to a enough shielding for the UV radiation 
by the dust absorption and the gas can be cooled to form the molecular hydrogen.
Otherwise, some of the detected molecular gas may be explained by the external origin, 
namely, the cold gas accretion into the galaxy \citep{you14}. 
\citet{you11} also found that the molecular gas is detected preferentially in the 
fast rotator early-type galaxies, while there is few slow rotators with the CO detection. 
Therefore the mass assembly history may be related with the existence of the molecular gas 
in early-type galaxies. Since the surface stellar mass density of the fast rotators 
tends to be lower than that of the slow rotators \citep{cap13}, the dust temperature 
may cause the difference between these two populations in the CO detection rate.

It is noted that low-mass passively evolving galaxies have the similar surface mass 
density with star-forming galaxies, and therefore their dust temperature is expected 
to be correspondingly low (see Figure \ref{fig:smdt}). The passive evolution of 
these low-mass galaxies may be explained by other mechanisms such as environmental 
or satellite quenching (e.g., \citealp{pen12}).

While we assumed the SED of the old stellar population to investigate the effect of 
the radiation from low-mass stars, there are young massive stars in star-forming galaxies. 
Although these massive stars are much luminous than low-mass stars and heat the surrounding 
dust, the ultraviolet light from these stars is efficiently absorbed by the surrounding 
dust and molecular hydrogen (self-shielding).
Therefore the temperature of the cold dust in disks of spiral galaxies tends to be 
relatively low ($T_{\rm dust} \sim 17$  K; \citealp{tab10}; \citealp{cle13}), 
while there is also the warm dust component. 
On the other hand, such self-shielding effect for the near-infrared radiation from 
low-mass stars is expected to be much lower. 
Thus the radiation from low-mass stars may play a role to heat the dust globally 
in star-forming galaxies when low-mass stars are sufficiently accumulated 
as the star formation proceeds.

The dust heating by low-mass stars can contribute to the passive evolution of 
massive compact galaxies found at $z\sim2$ (e.g., \citealp{tru07}; \citealp{van14}).
Since a very strong stellar radiation field is expected within such compact massive galaxies, 
the dust temperature is probably high in these objects. 
The effect of the high dust temperature may also be consistent with the 
inside-out scenario of the massive galaxy formation (e.g., \citealp{van10}; 
\citealp{pat13}). 
The star formation outside a compact core could be permitted if the radiation field 
is relatively low at outer regions, 
while the formation of molecular hydrogen is strongly suppressed within the core.
The direct measurements of the dust temperature and its profile  
for massive galaxies at high redshift by ALMA will be important to verify these scenarios.


\vspace{1pc}
We would like to thank the referee for many invaluable suggestions.
This work is based on observations taken by the 3D-HST Treasury Program (GO 12177 and 12328) 
with the NASA/ESA HST, which is operated by the Association of Universities for Research 
in Astronomy, Inc., under NASA contract NAS5-26555.  
Funding for SDSS-III has been provided by the Alfred P. Sloan Foundation, 
the Participating Institutions, the National Science Foundation, 
and the U.S. Department of Energy Office of Science. The SDSS-III web site is 
http://www.sdss3.org/.
SDSS-III is managed by the Astrophysical Research Consortium for the 
Participating Institutions of the SDSS-III Collaboration including the University of Arizona, 
the Brazilian Participation Group, Brookhaven National Laboratory, Carnegie Mellon University, University of Florida, the French Participation Group, the German Participation Group, 
Harvard University, the Instituto de Astrofisica de Canarias, 
the Michigan State/Notre Dame/JINA Participation Group, Johns Hopkins University, 
Lawrence Berkeley National Laboratory, Max Planck Institute for Astrophysics, 
Max Planck Institute for Extraterrestrial Physics, New Mexico State University, 
New York University, Ohio State University, Pennsylvania State University, 
University of Portsmouth, Princeton University, the Spanish Participation Group, 
University of Tokyo, University of Utah, Vanderbilt University, University of Virginia, 
University of Washington, and Yale University.
This publication makes use of data products from the Two Micron All Sky Survey, 
which is a joint project of the University of Massachusetts and 
the Infrared Processing and Analysis Center/California Institute of Technology, 
funded by the National Aeronautics and Space Administration and 
the National Science Foundation.
Data analysis were in part carried out on common use data analysis computer system 
at the Astronomy Data Center, ADC, of the National Astronomical Observatory of Japan.
YT acknowledges the financial support from the Japan Society for
the Promotion of Science (No. 23244031).

{}
\end{document}